\documentclass{sf2a-conf}
\usepackage{graphicx}

\begin{document}
\TitreGlobal{SF2A 2008}
\title{A Global Energetic Model for Microquasars: preliminary results and spectral energy distributions}
\author{Foellmi, C.} \address{Laboratoire d'Astrophysique de Grenoble, OSUG, U. Joseph Fourier, CNRS UMR 5571, 38400 Saint-Martin d'H\`eres, France}
\author{Petrucci, P.-O.$^1$} 
\author{Ferreira, J.$^1$} 
\author{Henri, G.$^1$} 
\author{Boutelier, T.$^1$} 
\runningtitle{Theoretical SEDs for microquasars}
\setcounter{page}{237} 


\maketitle
\begin{abstract}
We present preliminary results and observables from a model of microquasar based on a theoretical framework where stationary, powerful, compact jets are launched and then accelerated from an inner magnetized disk. This model aim at providing a consistent picture of microquasars in all their spectral states. It is composed of an outer standard accretion disk down to a variable transition radius where it changes to a magnetized disk, called the Jet Emitting Disk (JED). The theoretical framework providing the heating, we solve the radiative equilibrium and obtain the JED structure. Our JED solutions are rich, and reproduce the already known scheme where a cold optically-thick and a hot optically-thin solutions bracket a thermally unstable one. We present the model and preliminary results, whith a first attempt at reproducing the observed SED of XTE~J1118+480.
\end{abstract}
%
\section{Introduction}

Microquasars are stellar binaries with a short period in which one of the two components is a stellar-mass black-hole. The secondary component is a normal star filling its Roche lobe, and loosing mass through the first Lagrangian point. The ejected matter organizes itself in an accretion disk, from which jets are launched. Microquasar are one piece of a chain of objects with Active Galactic Nuclei and Gamma Ray Bursts, sharing the same ingredients and physics (Mirabel 2004).

Various theoretical works have been made to model the accretion disk, with the goal of reproducing the observations. The first kind of observations that is attempted to be reproduced is the spectral energy distribution (SED), which results from the combination of model {\it components} and the consistency with which they are combined, and the physical radiative processes occuring therein. Other observables, such as for example time-lag observed in X-rays or Quasi-Periodic Oscillations, are usually being modeled with more or less dedicated work, and reflects more the behavior of a given component (the disk for QPOs, e.g. Tagger \& Varni\`ere 2006, and the jet for time-lags, see for instance Kylafis et al. 2008).

Our own approach is the following: based on a {\it consistent} dynamical framework describing how jets are launched from a disk (see Ferreira 1997 and references therein), we build a microquasar model comprising an outer standard accretion disk down to a transition radius, and an inner disk from which jets are lauched, called the Jet Emitting Disk (JED). This model is a global attempt at explaining the general behavior of microquasars in all their spectral states (see Ferreira et al. 2006, Petrucci et al. 2008). Our goal here is to explore the richness of the solutions given by the radiative equilibrium of the JED, and produce theoretical SEDs that can be tested against observations.

\section{The model}

Our model for microquasars is based on the theoretical framework described in Ferreira et al. (2006, see in particular their Fig.~1). We model the outer standard accretion disk, from the transition radius $r_j$ onwards, with a multi-temperature blackbody spectrum, following Makishima et al. (1996). In the inner JED, comprised between the last stable orbit $r_{\textrm{\scriptsize{ISCO}}} = 6 R_{g}$ and $r_{j}$, we consider a one-temperature plasma with thermal particules only, and follow Mahadevan (1997) for the Bremsstrahlung,  Esin et al. (1996) for the Synchrotron.

We approximate the Compton cooling by using the parameter $\eta$ of Dermer et al. (1991), which gives the mean gain of energy in a Compton diffusion. We then solve the conservation of energy and the conservation of photons, assuming that in all cases, the compton intensity spectrum follows a power-law with a Wien bump at high energy. In order to obtain the three parameters of the comptonized spectrum, we link the Wien's bump normalisation $\gamma$ and the power-law index $\alpha$ (see Wardzinski \& Zdziarski, 2000, Equ.~23).

In our computations we consider two sources of photons that are being comptonized: the Synchrotron self-Compton on the inner JED photons (internal Compton), and the external Compton on photons coming from the standard accretion disk, and reaching the JED. This later contribution appeared to be of central importance in some cases where the transition radius $r_j$. The correct inclusion of these photons for the comptonization deserves a full and detailed geometrical treatment. As for now, we have simply considered that half of the photons coming from the inner radius of the standard accretion disk (i.e. at $r_j$) only are seen in the JED for comptonization by hot electrons.

\section{Radiative equilibrium of the disk and the JED solution}

\begin{figure}[!ht]
\begin{center}
\includegraphics[width=0.6\linewidth]{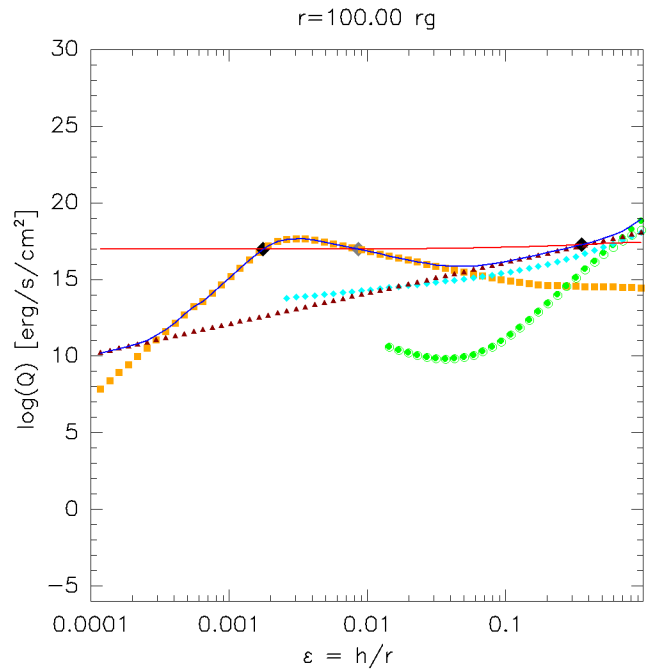}
\caption{The radiative equilibrium is computed here for the JED radius $r=100R_g$. The various cooling contributions are indicated by dots of various shapes. In orange, the Bremsstrahlung, cyan the external Compton, green the SSC, brown the advection. The red and blue lines indicate the heating and the total cooling respectively.}
\label{fig1}
\end{center}
\end{figure}

In order to compute theoretical SEDs, we need to compute the radiative equilibrium of the disk. We make the hypothesis, a posteriori verified, that the total pressure is dominated by the gaz pressure. Moreover, the vertical equilibrium of the disk links the temperature $T_e$ to the disk scale height $\epsilon \equiv h/r$. Consequently, the cooling is only a function of $\epsilon$. We compute the cooling by using the individual SEDs of each independant radius of the disk, and adding an advection term (which becomes significant only at $\epsilon \sim 0.1$). The heating of the disk is also a function of $\epsilon$ only, as provided by the theoretical framework (see Combet \& Ferreira 2008). We then solve the equation $Q^{-}(\epsilon) = Q^{+}(\epsilon)$ (see Fig~\ref{fig1}), where $Q^{-}(\epsilon) = Q_{\textrm{\scriptsize{Bremsstrahlung}}} + Q_{\textrm{\scriptsize{Synchrotron Compton}}} + Q_{\textrm{\scriptsize{External Compton}}} + Q_{\textrm{\scriptsize{Advection}}}$.

\begin{figure}[!h]
\begin{center}
\includegraphics[width=0.5\linewidth]{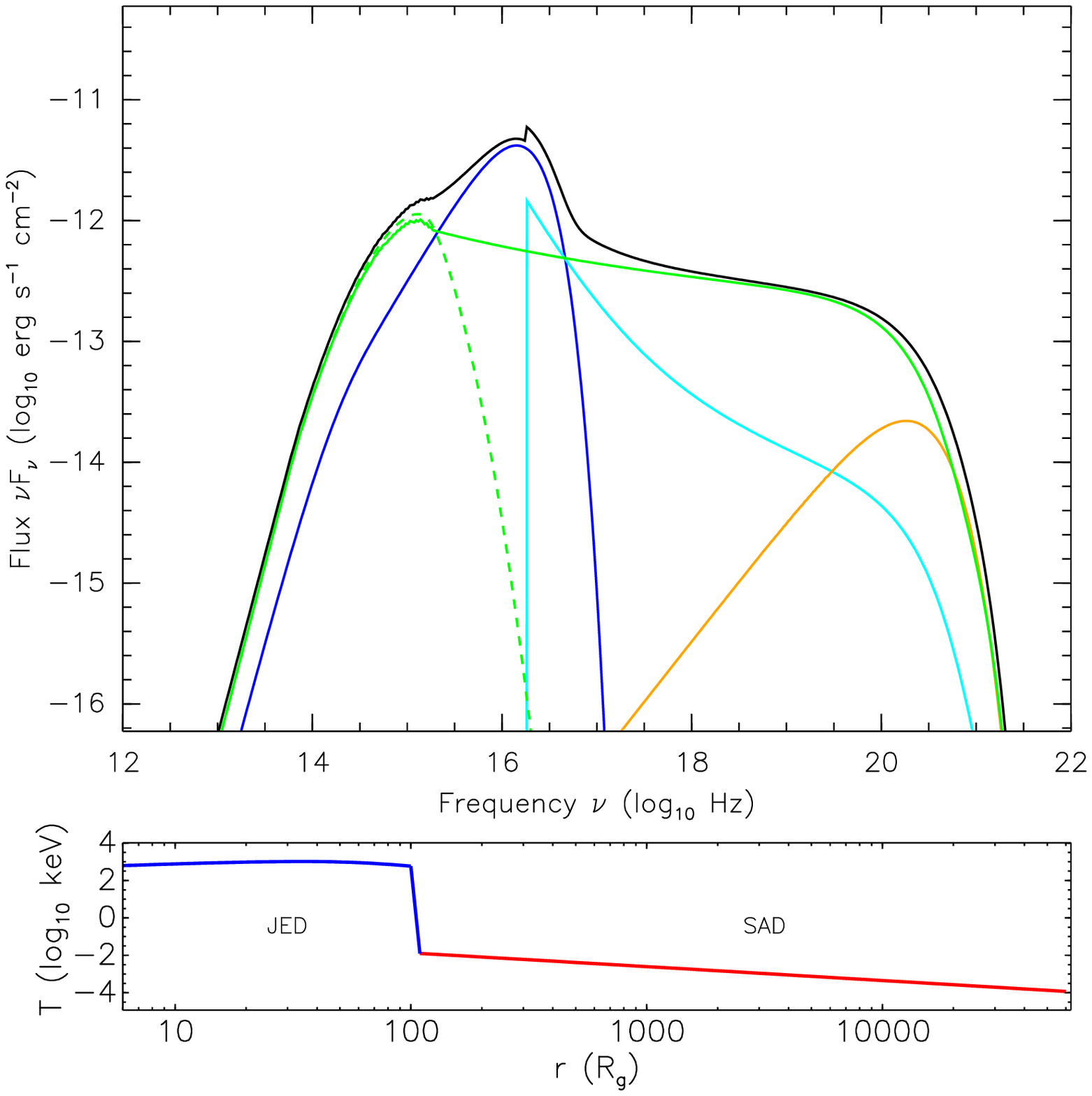}\includegraphics[width=0.5\linewidth]{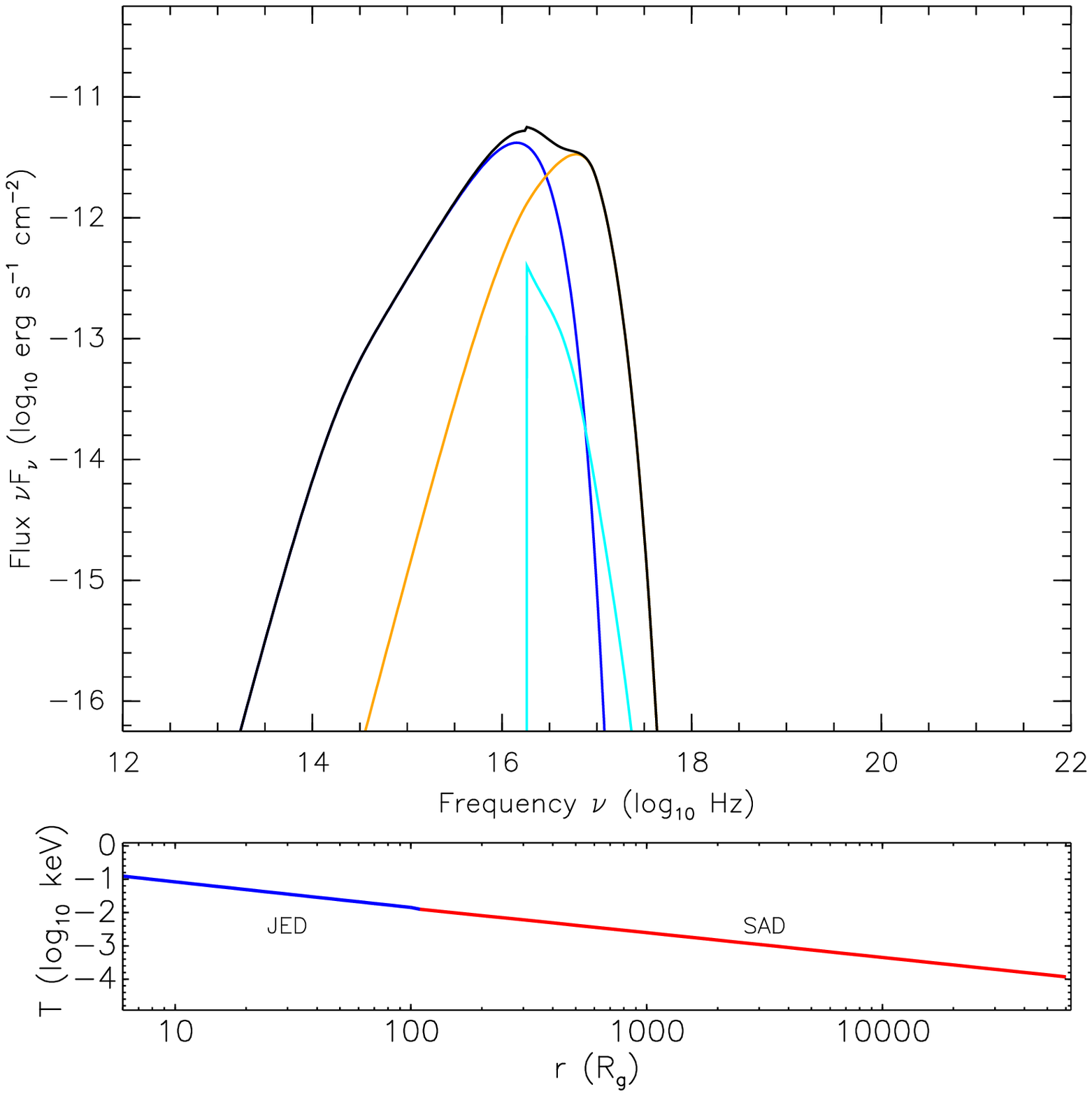}
\caption{Example of theoretical SEDs. Left: using the "hot" stable solution. Right: using the "cold" stable solution. The input parameters are identical to both figures: an accretion rate $\dot{M} = 10^{-2} M_{\textrm{\scriptsize{Edd}}} = 10^{-2} L_{\textrm{\scriptsize{Edd}}}/c^2$, an internal disk radius $r_i = 6 R_g$, a transition radius between the standard accretion disk and the JED $r_j = 100 R_g$, a magnetization $\mu = 1.0$, and the sonic Mach number $m_s = 1.0$, for an object with $M_{bh} = 10 M_{\odot}$, an inclination angle $i = 70^{\circ}$ and a distance of 10 kpc. The black line indicate our final SED, the jet is in red, the Bremsstrahlung in orange, the standard accretion disk in blue, the external Compton in cyan. The synchrotron is indicated in dashed green (it is not counted in the final SED), while the SSC is shown in plain green.}
\label{fig2}
\end{center}
\end{figure}

\begin{figure}[!t]
\begin{center}
\includegraphics[width=0.6\linewidth]{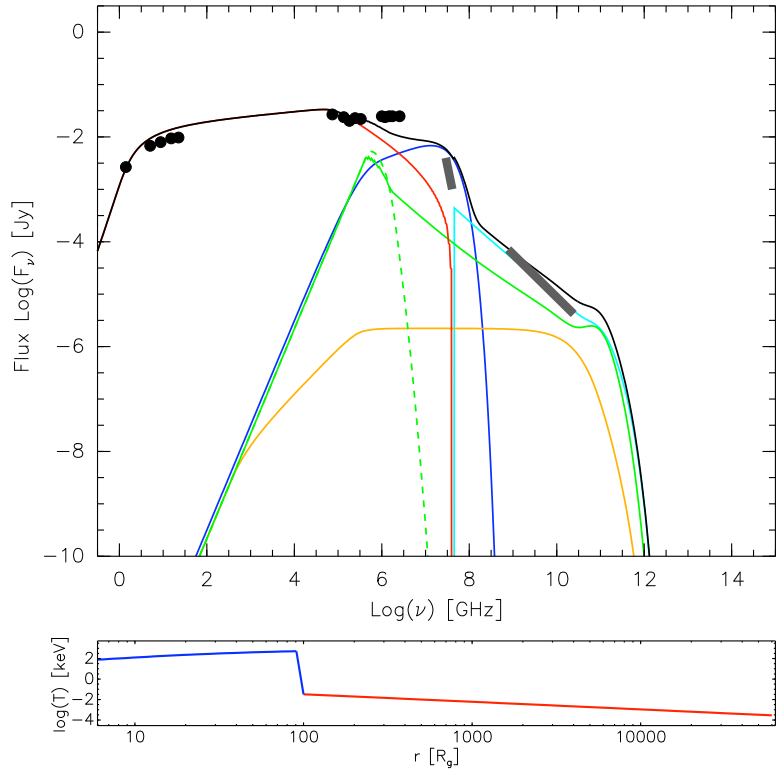}
\caption{Manual fit of the observed SED of XTE~J1118+480 taken from Markoff et al. (2001). The bottom panel shows the corresponding disk where the red line is the standard accretion disk, and in blue the hot JED disk. Parameters are: $M_{bh} = 6. M_{\odot}$, $\dot{M} = 0.2 M_{\textrm{\scriptsize{Edd}}}$ and $D = 1.8$ kpc taken from Markoff et al. (2001) and a hot JED disk with $r_j = 100 R_g$. The jet synchrotron model follows Heinz \& Sunyaev (2003).}
\label{fig3}
\end{center}
\end{figure}

Interestingly, our novel accretion disk solution often leads to three solutions of the thermal equilibrium: a cold optically-thick and a hot optically-thin thermally stable ones, and one middle thermally unstable solution (see black losanges in Fig.~\ref{fig1}). One interesting output of our model is that we are able to produce a hard X-ray spectrum even with thermal particles, and with the clear advantage of having consistently taken into account the power extracted  for the jet. This is illustrated in Fig.~\ref{fig2}. However, the jet is weakly radiative, and radiate through an unknown distribution of non-thermal particles. We thus model the jet emission by a pure synchrotron emission model, following Heinz \& Sunyaev (2003). We have then attempted to manually reproduce the observed SED of the galactic microquasar XTE~J1118+480 (see Markoff et al. 2001), as shown in Fig.~\ref{fig3}.

\section{Perspectives}

At that stage, our model can be tested against observational SEDs when data is available over a large frequency domain (as shown above), but mostly when the object is in the so-called low/hard state. Preliminary tests have shown that we cannot reproduce the high/soft state because of the absence in our model of a component that can produce the hard-tail power-law observed during these stages. In the future, we will tackle this issue along with the developement of a more realistic jet emission model. Our final goal is to be able to explore the hardness-intensity diagram and reproduce an hysteresis pattern.


\end{document}